\documentclass[12pt]{article}
\usepackage{latexsym,amsfonts,amssymb}

\begin{document}

\begin{center}
{\LARGE \bf Magnetic influence on the frequency of the soft-phonon mode in the incipient ferroelectric EuTiO$_3$}

\vskip 0.2cm

{\large  Qing Jiang$^{*}$ and Hua Wu}
\vskip 0.2cm 
{Department of Physics, Suzhou University, Suzhou, 215006, China}
\end{center}

\begin{abstract}
\quad 
The dielectric constant of the incipient ferroelectric EuTiO$_3$ exhibits a sharp decrease at about $5.5K$, at which temperature antiferromagnetic ordering of the Eu spins simultaneously appears, indicating coupling between the magnetism and dielectric properties. This may be attributed to the modification of the soft-phonon mode, $T_{1\mu }$, which is the main contribution to the large dielectric constant, by the Eu spins(7$\mu _B$ per Eu). By adding the coupling term between the magnetic and electrical subsystems as $
-g\sum\limits_l {\sum\limits_{\left\langle {i,j} \right\rangle } {q_l^2 } } \overrightarrow {S_i }  \cdot \overrightarrow {S_j } 
$ we show that the variation of the frequency of soft-phonon mode depends on the spin correlation between the nearest neighbors Eu spins and is substantially changed under a magnetic field. 
\\
\\
PACS number(s): 75. 50. Ee, 75. 80. +q
\\
\\
\\
\\
\\
\\
\vspace{3.0 cm}
\\
{*}Mailing address in China
\\
E-mail:qjiang@suda.edu.cn
\end{abstract} 

\newpage
\quad

{\bf{I. INTRODUCTION}}

The coupling between magnetism and transport properties in metals produces various interesting phenomena such as colossal magnetoresistance in perovskite manganites. Hund's rules dominate in this class of compounds, where the localized spins in the $t_{2g}$ state strongly couple with the itinerant electrons in the $e_g$ state. If a magnetic field is applied under such condition, the configuration of the localized spins changes, giving rise to a large magnetoresistance. Correspondingly, it may be desirable that coupling between magnetism and dielectric properties exist in certain magnetic insulators where the configuration of the localized spins to some extent affects the dielectric properties. A good example is yttrium magnanite YMnO$_3$. An inverse s-shaped anomaly in both dielectric constant and loss tangent has been detected near its magnetic ordering temperature, due to the frustration on the triangular lattice.$^{1, 2, 3}$ EuTiO$_3$ is another good example. Magnetic measurements show that EuTiO$_3$ is one of the few antiferromagnets with a positive Curie-Weiss temperature.$^{4}$ It has been reported$^{5}$ that the dielectric constant exhibits a sharp decrease at its $N\acute {e}el$ temperature $5.5K$. The crystal structure of EuTiO$_3$ is presented in Fig.1, which comes from the results presented in Ref.5. It exhibits a typical structure of perovskite titanates, where Eu ions are responsible for the antiferromagnetic ordering and its spin vector lies in the $ab$ plane while the atomic motion of the soft-phonon mode with the $T_{1u}$ symmetry is along the $c$ axis.$^{5}$ From the neutron-diffraction results EuTiO$_3$ has a type-G antiferromagnetic structure i.e., the six nearest-neighbor Eu ions to a given Eu ion have opposite spins while the twelve next-nearest-neighor Eu ions have parallel spins. In contrast to most of the perovskite titanates RTiO$_3$(R=rare earth), which usually bear trivalent R and Ti, EuTiO$_3$ has a magnetic divalent Eu because of the stable $4f^7$ electron configuration and accordingly a tetravalent Ti ($3d^0$).  

X-ray absorption near edge structure$^{6}$ has shown the absence of lattice distortion from the ideal perovskite structure in EuTiO$_3$ as well as in SrTiO$_3$, which is an incipient ferroelectric. EuTiO$_3$ is very similar to SrTiO$_3$ as an incipient ferroelectric, demonstrating no frequency dispersion, which is verified through dielectric measurements.$^{5}$ Incipient ferroelectrics are quite different from normal ferroelectrics as far as their dielectric properties are concerned. The dielectric constant increases with decreasing temperature, gradually deviating from the Curie-Weiss law, and saturates at a very low temperature. From the soft mode point of view, a normal incipient ferroelectric has a polar soft mode, but this polar soft mode does not freeze even in the neighbourhood of $0K$ because of the stabilization of the paraelectric phase by the quantum fluctuations.$^{7}$ EuTiO$_3$, however, behaves differently at low temperatures. Its dielectric constant shows a sharp decrease at its $N\acute {e}el$ temperature $5.5K$. Furthermore the experimental dielectric data of EuTiO$_3$ shows good agreement between the behavior of the dielectric constant and the calculated spin correlation $\left\langle {\vec S_i  \cdot \vec S_j } \right\rangle$ between nearest neighbors Eu spins under different magnetic fields. It is enlightening to consider the possible relations between the magnetic and electrical subsystems and we should therefore take the coupling interaction between the magnetic and electrical subsystem into account in such systems at low temperatures. As a matter of fact, for EuTiO$_3$ the superexchange between the nearest neighbours takes place through the mediation of the O $2p$ state. Moreover the O ion plays a role in the soft-phonon mode with $
T_{1\mu } 
$ symmetry
in the center of the Brillouin zone. Therefore the variation of the Eu spin configuration may affect the behaviour of the soft-phonon mode.  Furthermore it is clear that the soft-phonon mode with low frequency is responsible for the dielectric properties via the empirical formula $
\chi _{\vec q}  \approx \displaystyle\frac{1}{{m(\omega _{\vec q}^2  - \omega ^2 )}}
$
 in
EuTiO$_3$. Therefore a plausible speculation is that the variation of the spin configuration at Eu site near and below $T_N$ will affect the behavior of the soft-phonon mode and lead to special dielectric properties.

  X. S. Gao et al.$^{8}$ utilized Monte Carlo simulations on the basis of the Ising-DIFFOUR model$^{9}$ to investigate the phase transition in the two dimensional ferroelectromagnetic lattice where the spin moment and polarization interact. For EuTiO$_3$, however, things are different. First EuTiO$_3$ is  a typical example where the Heisenberg model applies instead of the Ising model, since the magnetic ions are in a s-type orbital and the exchange interaction is isotropic. Secondly, based on available experimental data for EuTiO$_3$, Katsufuji and Takagi$^{5}$ proposed that the dielectric constant and spin correlation $\left\langle {\vec S_i  \cdot \vec S_j } \right\rangle$ are related as $
\varepsilon  = \varepsilon _0 \left( {1 + \alpha \left\langle {\vec S_i  \cdot \vec S_j } \right\rangle } \right)
$, which differs from the coupling term previously suggested in Ref.8. Thus it is reasonable to add a term in the Hamiltonian of the form $
g\sum\limits_l {\sum\limits_{\left\langle {i,j} \right\rangle } {q_l^2 } } \overrightarrow {S_i }  \cdot \overrightarrow {S_j } 
$ to describe  the coupling interaction between the electrical and magnetic subsystems. On such an assumption, we obtain analytical relations between the soft-phonon mode and the spin correlation as well as between the dielectric constant and the spin correlation, which are in good agreement with the experimental results. Our results show that the squared frequency of the soft-phonon mode is related to the spin correlation $\left\langle {\vec S_i  \cdot \vec S_j } \right\rangle$. We also discuss the behaviour of the soft-phonon mode at different temperatures and under different magnetic fields.

In our present paper we first deal with the electrical subsystem, including electrical-magnetic coupling, using the perturbation theory. In order to explore the  soft-phonon behavior, we apply the soft mode theory under the mean-field approximation. Secondly, to further investigate the temperature and magnetic field dependence of the frequency of the soft-phonon mode, we study the magnetic spin correlation, which influences the behavior of the soft-phonon mode. Finally we give our numerical results and detailed discussion.\\

 {\bf{II. MODELS AND ANALYSIS}}

We consider the Hamiltonian for a three dimensional cubic system with periodic boundary conditions. Two parameters, $q_l$ and $S_i$  are proposed here to represent the electrical displacement at Ti site and the Heisenberg spin at Eu site, respectively. In addition, the coupling interaction between the electrical and magnetic subsystems is taken into account. Following Janssen's model$^{9}$ the Hamiltonian for this system can be presented as following:\begin{equation}
H = H^e  + H^m  + H^{me}, 
\end{equation} where $H^e$ denotes the Hamiltonian for the electrical subsystem, $H^m$ is the Hamiltonian for the magnetic subsystem and $H^{me}$ the coupling interaction between the two subsystems. Here the soft mode theory under the mean-field approximation is applied to solve $H^e$. In treating $H^e$, the Eu and O ions are fixed and only Ti ions vibrate as was supposed by Slater.$^{10}$ According to the soft mode theory, the atoms vibrate like a hamonic oscillator with small additional anharmonic part. That is to say $q_l$ at each site is subjected to a single-well potential. Thus $H^e$ in the absence of electric field can be written as following:\begin{equation}
H^e  = \sum\limits_l {\left[ {\displaystyle\frac{{p_l^2 }}{{2m}} + V\left( {q_l } \right)} \right]}  - \displaystyle\frac{1}{2}\sum\limits_l {\sum\limits_{l'} {\nu _{ll'} } } q_l q_{l'}, 
\end{equation} where $p_l$ is the momentum at site $l$, ${V\left( {q_l } \right)}$ is the localized potential function and $
\displaystyle\frac{1}{2}\sum\limits_l {\sum\limits_{l'} {\nu _{ll'} } } q_l q_{l'} 
$ denotes the two-body interaction potential. We use the simplest form to describe the anharmonic motion at Ti site. Thus \begin{equation}
V(q_l ) = \displaystyle\frac{1}{2}m\Omega _0^2 q_l^2  + \frac{1}{4}\gamma q_l^4,
\end{equation}
 where $
\Omega _0 
$
 is the inherent frequency, $
\displaystyle\frac{1}{2}m\Omega _0^2 q_l^2 
$
 is the harmonic part and $\displaystyle\frac{1}{4}\gamma q_l^4 
$ the anharmonic one. 

The interaction between Eu $4f$ spins originates from (i)superexchange(antiferromagnetic) through the O $2p$ state and also from (ii)indirect exchange through the Eu $5d$ state. So $H^m$ with $
S = \displaystyle\frac{7}{2}
$ can be written as :
\begin{equation}
H^m  = \sum\limits_{\left\langle {i,j} \right\rangle } {J_1 } \vec S_i  \cdot \vec S_j  + \sum\limits_{[i,j]} {J_2 } \vec S_i  \cdot \vec S_j  - \sum\limits_i {\vec h \cdot \vec S_i }, 
\end{equation}
 where $S_i$ is the Heisenberg spin at Eu site $i$, and the exchange integrals $J_1  = 0.037k_B$ $K$ and $J_2  = -0.069k_B$ $K$ to produce $
T_N  = 5.5K
$
represent the antiferromagnetic coupling between the nearest neighbours and the ferromagnetic coupling between the next nearest neighbours, respectively. Here $h$ is the external magnetic field parallel to the $z$ axis. $
\left[ {i,j} \right]
$
and $
\left\langle {i,j} \right\rangle 
$
 denote the once summation over the next nearest neighbors and the nearest neighbors, respectively. 

As we pointed out in the previous section we propose the coupling term between the magnetic and electrical subsystems as:
\begin{equation}
H^{me}  =  - g\sum\limits_l {\sum\limits_{\left\langle {i,j} \right\rangle } {q_l^2 } } \overrightarrow {S_i }  \cdot \overrightarrow {S_j },
\end{equation} where $g$ is the coupling coefficient. Here only the quadratic term of $q_l$ is taken because of the time reversal and space reversal symmetries. Here $l$ represents the lattice site of Ti ion, $i$ and $j$ denote distinct nearest neighbor Eu ions to a given Ti ion.

First we deal with the electrical subsystem, considering electrical-magnetic coupling. Under the mean-field approximation, the single-ion Hamiltonian can be written as:
\begin{equation}
H_l^E  = H_l^e  + H_l^{me}  = \frac{{p_l^2 }}{{2m}} + (\frac{1}{2}m\Omega _0^2  - z_1 g\left\langle {\overrightarrow {S_i }  \cdot \overrightarrow {S_j } } \right\rangle )q_l^2  + \frac{1}{4}\gamma q_l^4  - \sum\limits_{l'} {\nu _{ll'} \left\langle {q_{l'} } \right\rangle q_l }, 
\end{equation}
 where $z_1  = 12$
 is the number of the spin correlation that will directly affect the motion of Ti in one unit cell and $\displaystyle\frac{1}{4}\gamma q_l^4 
$ can be regarded as the small perturbation part. To facilitate the further discussion, we define $b = \displaystyle\frac{{24g}}{{m\Omega _0^2 }}\left\langle {\overrightarrow {S_i }  \cdot \overrightarrow {S_j } } \right\rangle$ and $\Omega _0 (g) = \Omega _0 \sqrt {1 - b}$. The magnetic effect on the electrical subsystem is embodied by $
\Omega _0 (g)
$. Thus \begin{equation}
H_l^E  = \frac{{p_l^2 }}{{2m}} + \frac{1}{2}m\Omega _0^2 (g)q_l^2  + \frac{1}{4}\gamma q_l^4  - \sum\limits_{l'} {\nu _{ll'} \left\langle {q_{l'} } \right\rangle q_l }. 
\end{equation}
The frequency of the soft-phonon mode is given through soft-mode theory:$^{11}$ \begin{equation}
m\omega _{\vec q}^2 (g) = m\Omega _0^2 (g) + 3\gamma \sigma _T  - \nu _{\vec q}, 
\end{equation} where \begin{equation}
\nu _{\vec q}  = \sum\limits_{l'} {\nu _{ll'} \exp \left[ { - i\vec q \cdot (\vec R_l  - \vec R_{l'} )} \right]},
\end{equation} is the interaction potential at the wave vector $
{\vec q}$. The key point is how to calculate the fluctuation of displacement $
\sigma _T  = \left\langle {q_l  - \left\langle {q_l } \right\rangle } \right\rangle ^2 
$. In incipient ferroelectrics, quantum effects are very important at very low temperatures, so quantum statistics must be adopted. At the same time we notice that no ferroelectric phase transition takes place in the incipient ferroelectrics and the average displacement is zero. 
Thus we use quantum statistics and perturbation theory to obtain the total displacement fluctuation at very low temperatures, which contains both quantum fluctuations and thermal fluctuations:
\begin{equation}
\sigma _T  = \left\langle {q_l  - \left\langle {q_l } \right\rangle } \right\rangle ^2  = \displaystyle\frac{\hbar }{{2m\Omega _0 (g)}}coth(\displaystyle\frac{{\hbar \Omega _0 (g)}}{{2k_B T}}) - \displaystyle\frac{{3\gamma }}{{16m^3 \Omega _0^4 (g)}}\left[ {coth^2 (\displaystyle\frac{{\hbar \Omega _0 (g)}}{{2k_B T}}) + 1} \right],
\end{equation}
For EuTiO$_3$ the soft-phonon mode is in the center of Brillouin zone($\vec q = 0
$), therefore according to the single mode theory under the mean-field approximation the frequency of the soft-phonon mode can be written:
 \begin{equation}
m\omega _{\vec q = 0}^2 (g) = m\Omega _0^2 (g) + 3\gamma \sigma _T  - \nu _{\vec q = 0}  = \frac{{3\gamma }}{{m\Omega _0^2 (g)}}\left[ {\frac{{\hbar \Omega _0 (g)}}{2}coth(\frac{{\hbar \Omega _0 (g)}}{{2k_B T}}) - k_B T_0 } \right],
\end{equation}
 where \begin{equation} 
k_B T_0  = \frac{{m\Omega _0^2 (g)}}{{3\gamma }}\left[ {\nu _{\vec q = 0}  - m\Omega _0^2 (g)} \right].
\end{equation}
 Obviously if the coupling effect is not included, i.e., $
g = 0
$, Eq. (11) is just the expression of the frequency of the soft-phonon mode for normal incipient ferroelectrics. The frequency of the soft-phonon mode does not go to zero even close to $0K$.  
In order to directly exhibit how the spin correlation has effect on the behavior of the soft-phonon mode we arrange Eq. (11) as following:
\begin{equation}
m\omega _{\vec q = 0}^2 (g) \approx m\omega _{\vec q = 0}^2 (0)(1 - \alpha \left\langle {\overrightarrow {S_i }  \cdot \overrightarrow {S_j } } \right\rangle ),
\end{equation}
where \begin{equation}
\alpha  = \displaystyle\frac{{24g}}{{m^2 \Omega _0^2 \omega _{\vec q = 0}^2 (0)}}\left[ {m\Omega _0^2  - \frac{{3\gamma \hbar }}{{4m\Omega _0 }}coth\left( {\displaystyle\frac{{\hbar \Omega _0 }}{{2k_B T}}} \right) + \displaystyle\frac{{\hbar \Omega _0 }}{{2k_B Tsinh^2 \left( {\displaystyle\frac{{\hbar \Omega _0 }}{{2k_B T}}} \right)}}} \right].
\end{equation}
Here $\omega _{\vec q = 0} (0)$
 is the frequency of the soft-phonon mode as the coupling interaction is not considered and it is temperature independent at very low temperatures. Because the coupling interaction only exists near the $N\acute {e}el$ ordering temperature, $\alpha$ can be approximately written as \begin{equation}
\alpha  = \frac{{24g}}{{m^2 \Omega _0^2 \omega _{\vec q = 0}^2 (0)}}\left[ {m\Omega _0^2  - \frac{{3\gamma \hbar }}{{4m\Omega _0 }}} \right].
\end{equation} The static susceptibility, which embodies the dielectric constant, is given by:
\begin{equation}
\chi  = \frac{1}{{m\omega _{\vec q = 0}^2 (0)}}(1 + \alpha \left\langle {\overrightarrow {S_i }  \cdot \overrightarrow {S_j } } \right\rangle ) = \chi (0)(1 + \alpha \left\langle {\overrightarrow {S_i }  \cdot \overrightarrow {S_j } } \right\rangle )
\end{equation} where $
\chi (0)
$ is the static susceptibility when the coupling interaction is not included. Thus the dielectric constant can be derived from obove:\begin{equation}
\varepsilon  = \varepsilon (0)(1 + \beta \left\langle {\overrightarrow {S_i }  \cdot \overrightarrow {S_j } } \right\rangle ),
\end{equation} where \begin{equation}
\varepsilon (0) = A + \chi (0),
\end{equation} and \begin{equation}
\beta  = \displaystyle\frac{{\chi (0)\alpha }}{{A + \chi (0)}}.
\end{equation}
Here $A$ is the background dielectric constant. Eq. (17) is in good agreement with the dielectric properties fitted from the experimental data$^{5}$.

It is easy to show that there exists immediate relations between the behaviour of the soft-phonon  mode and spin configuration. To investigate how the frequency of the soft-phonon mode varies under different conditions of spin correlation, we take into account the spin correlation in the coupling system. Then Hamiltonian for the magnetic subsystem can be written:
\begin{equation}
H^M  = H^m  + H^{me}  = \sum\limits_{\left\langle {i,j} \right\rangle } {(J_1  - g\sum\limits_l {q_l^2 } )} \overrightarrow {S_i }  \cdot \overrightarrow {S_j }  + J_2 \sum\limits_{[i,j]} {\overrightarrow {S_i }  \cdot \overrightarrow {S_j } }  - \sum\limits_i \vec h  \cdot \overrightarrow {S_i }. 
\end{equation}
 This magnetic subsystem can be divided into two sublattices $a$ and $b$. Thus \begin{equation}
H^M  = H_a  + H_b. 
\end{equation}
where $H_a$ and $H_b$ are the Hamiltonians belong to $a$ and $b$ sublattices, respectively. Under the mean-field approximation \begin{equation}
H_a  = H_a^x \sum\limits_{ai} {S_{ai}^x }  + H_a^z \sum\limits_{ai} {S_{ai}^z }, 
\end{equation} and \begin{equation}
H_b  = H_b^x \sum\limits_{ai} {S_{bi}^x }  + H_b^z \sum\limits_{bi} {S_{bi}^z }, 
\end{equation} where \begin{equation}
H_{a(b)}^x  = 6(J_1  - z_2 g\left\langle {q_l^2 } \right\rangle )\left\langle {S_{b(a)}^x } \right\rangle  + 12J_2 \left\langle {S_{a(b)}^x } \right\rangle, 
\end{equation} \begin{equation}
H_{a(b)}^z  =  - h + 6(J_1  - z_2 g\left\langle {q_l^2 } \right\rangle )\left\langle {S_{b(a)}^z } \right\rangle  + 12J_2 \left\langle {S_{a(b)}^z } \right\rangle. 
\end{equation} Here $x$, $z$ denote the spin components and $
z_2  = 4
$ is the number of Ti that will directly affect the spin configuration at each pair of correlation. Thus the spin components of the different sublattices can be obtained: \begin{equation}
\left\langle {S_a^z } \right\rangle  = \displaystyle\frac{{ - H_a^z }}{{2\sqrt {\left( {H_a^x } \right)^2  + \left( {H_a^z } \right)^2 } }}\displaystyle\frac{{\sum\limits_{i = 1}^4 {(2i - 1)sinh} (\displaystyle\frac{{2i - 1}}{{2k_B T}}\sqrt {\left( {H_a^x } \right)^2  + \left( {H_a^z } \right)^2 } )}}{{\sum\limits_{i = 1}^4 {cosh} (\displaystyle\frac{{2i - 1}}{{2k_B T}}\sqrt {\left( {H_a^x } \right)^2  + \left( {H_a^z } \right)^2 } )}},
\end{equation}
\begin{equation}
\left\langle {S_b^z } \right\rangle  = \displaystyle\frac{{ - H_b^z }}{{2\sqrt {\left( {H_b^x } \right)^2  + \left( {H_b^z } \right)^2 } }}\displaystyle\frac{{\sum\limits_{i = 1}^4 {(2i - 1)sinh} (\displaystyle\frac{{2i - 1}}{{2k_B T}}\sqrt {A^2  + B^2 } )}}{{\sum\limits_{i = 1}^4 {cosh} (\displaystyle\frac{{2i - 1}}{{2k_B T}}\sqrt {A^2  + B^2 } )}}
\end{equation}
\begin{equation}
\left\langle {S_a^x } \right\rangle  = \displaystyle\frac{{H_a^x }}{{H_a^z }}\left\langle {S_a^z } \right\rangle,
\end{equation}
\begin{equation}
\left\langle {S_b^x } \right\rangle  = \displaystyle\frac{{H_b^x }}{{H_b^z }}\left\langle {S_b^z } \right\rangle.
\end{equation}
The spin correlation between the nearest neighbours can be approximately decoupled as:
\begin{equation}
\left\langle {\vec S_i  \cdot \vec S_j } \right\rangle  = \left\langle {S_a^x } \right\rangle \left\langle {S_b^x } \right\rangle  + \left\langle {S_a^z } \right\rangle \left\langle {S_b^z } \right\rangle. 
\end{equation}
For a complex system where coupling exists between the magnetic and electrical subsystems one should consider the subsystem results obtained by including the coupling term $
g\sum\limits_l {\sum\limits_{\left\langle {i,j} \right\rangle } {q_l^2 } } \overrightarrow {S_i }  \cdot \overrightarrow {S_j } 
$.\\

{\bf{III. RESULTS AND DISCUSSION}} 

Using Eq. (17)-Eq. (19) and Eq. (22)-Eq. (30) we fit the experimental results$^{5}$ in the absence of magnetic field. We not only get the parameter 
$T_0  =  - 25K
$, which is the same as that in Ref.5, but also we get the coupling coefficient $g$ as $9.42 \times 10^{ - 7}$, which is doubtlessly small but plays an important role on the behavior of the soft-phonon mode. The parameters $\Omega _0 
$ and $\gamma 
$ are also calculated as: $
\Omega _0  \approx 2.13 \times 10^{13} 
$ and $
\gamma  \approx 3.59 \times 10^{19} 
$. Now we use these parameters to study the effect of coupling on the frequency of the soft-phonon mode.

The spin correlation $
\left\langle {\vec S_i  \cdot \vec S_j } \right\rangle 
$ versus temperature under different magnetic field in the presence of  electrical-magnetic coupling is shown in Fig.2. Variation of the spin configuration can be realized by varying the temperature and the external magnetic field. The spin correlation does not change significantly if the coupling interaction is not neglected($g \to 0$). That is to say, the electrical effect is not strong enough to substantially reduce the spin correlation between the nearest neighbours.

But the magnetic influence on the behavior of the soft-phonon mode is rather evident. The temperature dependence of the frequency of the soft-phonon mode under different magnetic field is presented in Fig.3. We focus on the soft mode properties for $
T \le 30K
$, where the frequency of the soft-pnonon mode as well as the dielectric constant shows little temperature variation if the electrical-magnetic coupling is neglected($g = 0$; solid line). When the electrical-magntic coupling is included, the soft-phonon frequency increases remarkably below the antiferromagnetic ordering temperature (about $5.5K$ for $
h = 0
$), instead of the soft mode behavior of normal incipient ferroelectrics. One can infer that the antiferromagnetic ordering stiffens the soft phonon mode. Consequently, the dielectric constant decreases below the $N\acute {e}el$ temperature. However, when higher magnetic fields  are applied, sufficient to induce ferromagnetic ordering, softening of the soft-phonon mode can be observed. Such softening is responsible for the increase of the dielectric constant$^{5}$. To summarize, our results suggest that the soft-phonon mode is modified by the Eu spins. The configuration of Eu spins determines the hybridization between the Eu orbitals and the O $2p$ orbitals and then gives rise to the modification of the behaviour of the soft-phonon mode that contains the Ti - O stretching motion. 

The soft-phonon frequency versus applied magnetic field was also computed(Fig.4). We can see that for $
T \le 20K
$, the frequency of the soft-phonon mode decreases with increasing magnetic field, but the exact behavior of $
\omega (h)
$ depends on $T$. At $2K$, $
\omega (h)
$  decreases sharply as $h$ increases for $
h \le 1.5T
$. For $
h \ge 1.5T
$, $
\omega (h)
$ decreases more slowly as $h$ increases, owing to lesser variation of the spin correlation $\left\langle {\vec S_i  \cdot \vec S_j } \right\rangle$. For $
T = 4K
$, $
\omega (h)
$ shows less variation, however a kink in $
\omega (h)
$ still appears. At $5.5K$ the curve is more smooth and only a very slow decrease with the increasing magnetic field is observed. When a high magnetic field is applied, the spin correlation under the ferromagnetic ordering lowers the frequency. The lower the temperature the stronger the effect. For low magnetic field, however, the antiferromagnetic ordering dominates and the frequency of soft-phonon mode is enhanced relative to $
\omega (0)
$. The lower the temperature the greater the enhancement.  For $
T \ge 20K
$, at which the spin correlation becomes minuscule, $
\omega (h,T) 
$ is nearly independent of $h$ and increases as $T$ increases. This is the typical behavior for normal incipient ferroelectrics within the framework of classical statistics. At such high temperatures, the spin correlation is too weak to significantly affect the soft mode. Under such conditions the frequency of the soft-phonon mode in quantum statistics reduces to the classical Curie-Weiss result: $
\omega  \propto (T - T_0 )^{\frac{1}{2}} 
$
\\ 

{\bf{IV. SUMMARY:}}

EuTiO$_3$ is regarded as both a typical type-G antiferromagnet with $
S = \frac{7}{2}$
 at 
Eu$^{2 + }$ 
 site and an incipient ferroelectric with the perovskite crystal structure. This special property makes the frequency of the soft-phonon mode as well as the dielectric constant deviate from the behavior in nomal incipient ferroelectric, resulting from coupling between the magnetic and electrical subsystems. From our discussion above, variation of the spin correlation $
\left\langle {\overrightarrow {S_i }  \cdot \overrightarrow {S_j } } \right\rangle 
$
 causes a remarkable variation of the soft-phonon frequency, which is regarded as the main reason for the special dielectric properties of EuTiO$_3$.
 
\newpage
\begin{center} {REFERENCES} \end {center}

\par
$^{1}$Alessio Filippetti and Nicola A. Hill, Phys. Rev. B. {\bf 65} 195120 (2002).
\par
$^{2}$Z. J. Huang, Y. Cao, Y. Y. Sun, Y. Y. Xue and C. W. Chu, Phys. Rev. B. {\bf 56} 2623 (1997).
\par
$^{3}$T. Katsufuji, S. Mori, M. Masaki, Y. Moritomo, N. Yamamoto and  H. Takagi, Phys. Rev. B. {\bf 64}  104419 (2001).
\par
$^{4}$T. R. McGuire, M. W. Shafer, R. J. Joenk, H. A. Alperin, and S. J. Pickart, J. Appl. Phys {\bf 31} 981 (1966).
\par
$^{5}$T. Katsufuji and H. Takagi, Phys. Rev. B. {\bf 64} 054415 (2001).
\par
$^{6}$B. Ravel, E. A. Stern, Physica B. {\bf 208-209} 316 (1995).
\par
$^{7}$W. Zhong,  D. Vanderbilt, Phys. Rev. B {\bf 53} 5047 (1996).
\par
$^{8}$X. S. Gao, J. M. Liu, X. Y. Chen, and Z. G. Liu, J. Appl. Phys {\bf 88} 4250 (2000).
\par
$^{9}$T. Janssen and J. A. Tjion, Phys. Rev. B {\bf 24} 2245 (1981).
\par
$^{10}$J. C. Slater, Phys. Rev. {\bf 78} 748 (1950).
\par
$^{11}$K. A. M\"{u}ller, H. Burkard, Phys. Rev. B {\bf 19} 3593 (1979).

\newpage
\begin{center}{CAPTION OF FIGURES}\end{center}

Fig.1: The crystal structure of EuTiO$_3$. White circles and a black circle represents O and Ti, respectively. The white thick arrows are the antiferromagnetic ordering of Eu spins. The black thin arrows exhibit the atomic motion of the soft-phonon mode.\\

Fig.2: The temperature dependence of pair correlation $\left\langle {\overrightarrow {S_i }  \cdot \overrightarrow {S_j } } \right\rangle$
between the nearest neighbours at Eu$^{2 + }$. ($
g = 9.42 \times 10^{ - 7} 
$
)\\ 

Fig.3: The temperature dependence of the frequency of soft-phonon mode under different magnetic fields. ($
g = 9.42 \times 10^{ - 7} 
$
)\\

Fig4: The frequency of the soft-phonon mode versus magnetic field at different temperatures.
($
g = 9.42 \times 10^{ - 7} 
$
)\\
\end{document}